\documentclass[10pt]{iopart} %12pt or 10pt

\usepackage{iopams}
\usepackage{graphicx}% Include figure files
\usepackage{txfonts}

\begin{document}

\vspace*{-15mm}

\title[%
Anomalous magnetic interference of cross junctions]{%
Anomalous magnetic interference of cross-type Josephson junctions 
exposed to oblique magnetic fields
}

\author{Yasunori Mawatari}

\address{%
National Institute of Advanced Industrial Science and Technology (AIST), \\
Tsukuba, Ibaraki 305--8568, Japan
}
%\ead{y.mawatari@aist.go.jp}

% \vspace{10pt}
% \begin{indented}
% \item[]\today
% \end{indented}
%
\begin{abstract}
Gauge-invariant phase difference and critical currents of cross-type Josephson junctions with thin and narrow superconducting strips exposed to three-dimensional magnetic fields are theoretically investigated. 
When a sandwich-type Josephson junction in the $xy$ plane is exposed to parallel magnetic fields $H_x$ and $H_y$, the phase difference linearly depends on the spatial coordinates, $x$ and $y$, and the critical currents exhibit the standard Fraunhofer-type magnetic interference. 
The perpendicular field $H_z$, on the other hand, nonlinearly modulates the distribution of the phase difference and the critical currents as the functions of the oblique field exhibit anomalous magnetic interference. 
We obtain simple analytical expressions for critical currents of small cross-type junctions by 
neglecting the effects of self-field and trapped vortices. 
The resulting dc critical currents show anomalous and diverse interference patterns depending on the parallel and perpendicular magnetic fields. 
\end{abstract}

%\noindent{\it Keywords}: cross junction, critical current, magnetic interference
%\keywords{cross junction, critical current, magnetic interference}
%\submitto{\JPD}
% \maketitle
% For two-column output uncomment the next line and choose [10pt] 
% rather than [12pt] in the \documentclass declaration
%\ioptwocol %\small

\section{Introduction} %************************************************************
One of the most fascinating behaviors of the static response of Josephson junctions is the interference patterns in the magnetic-field dependence of the dc critical current $I_{\rm c}$. 
The $I_{\rm c}$ of small junctions exhibit Fraunhofer-type interference patterns~\cite{Rowell_63,Jaklevic_64,Josephson_65}, and the interference behavior in response to a magnetic flux in steps of the flux quantum $\phi_0=h/2e=2.07\times 10^{-15}$ Wb is a key phenomenon for application to ultrasensitive magnetometers, superconducting quantum interference devices (SQUIDs)~\cite{Jaklevic_64,Barone-Paterno_82,SQUID_04}. 

Anomalous magnetic interference in $I_{\rm c}$ of Josephson junctions has been extensively investigated by considering the effects of inhomogeneous current density in the junction~\cite{Barone-Paterno_82,Dynes_71}, trapped vortices~\cite{Miller_85,Gubankov_91,Golod_10,Clem_11}, local current injection~\cite{Nappi_02,Goldobin_04,Kogan_14}, ferromagnetic $\pi$ junctions~\cite{Kemmler_10,Alidoust_12,Borcsok_19}, and semiconductor junctions~\cite{Suominen_17}. 
Magnetic interference in $I_{\rm c}$ are also analyzed to investigate the edge states in topological insulators~\cite{Lee_14,Hart_14,Pribiag_15} and in graphene~\cite{Allen_16}. 

Here we demonstrate that anomalous interference patterns appear even in the naive case of sandwich-type Josephson junctions in \emph{oblique} magnetic fields. 
Although the response of junctions to \emph{parallel} magnetic fields is well understood~\cite{Barone-Paterno_82}, only a limited number of works have been published on the response to \emph{perpendicular} and/or \emph{oblique} magnetic fields~\cite{Miller_85,Rosenstein_75,Hebard_75,Monaco_08,Watanabe_05}. 
The systematic response of dc critical current of Josephson junctions to \emph{oblique} magnetic fields remains unclear. 

In this paper, we theoretically investigate the dc critical current $I_{\rm c}$ of cross-type junctions in oblique magnetic fields. 
We derive simple analytical expressions for $I_{\rm c}$ as a function of the three-dimensional magnetic field $\bi{H}=(H_x,H_y,H_z)$ by assuming that the magnetic screening effect in small junctions is weak and that no vortices are trapped in the junctions. 
We demonstrate a variety of anomalous interference patterns in the field dependence of $I_{\rm c}$, although the analytical formulae of $I_{\rm c}$ are quite simple. 

This paper is organized as follows. 
The static two-dimensional distribution of the gauge-invariant phase difference in small sandwich-type Josephson junctions is considered in Sec.~\ref{sec:basic-eqs}. 
Analytical expressions for $I_{\rm c}$ of cross-type Josephson junctions are derived and a variety of the interference patterns of $I_{\rm c}$ as functions of the parallel and perpendicular (i.e., in-plane and out-of-plane) magnetic fields are demonstrated in Sec.~\ref{sec:Ic-cross-jj}. 
Our results are summarized in Sec.~\ref{sec:summary}.

\section{Gauge-invariant phase difference 
\label{sec:basic-eqs}} %************************************************************
In this section we consider the static two-dimensional distribution of the gauge-invariant phase difference in sandwich-type Josephson junctions by neglecting the effects of self-field and trapped vortices. 

\subsection{Basic equations for two-dimensional Josephson junctions} %********************
\begin{figure}[t]%*************
	\center\includegraphics[width=80mm]{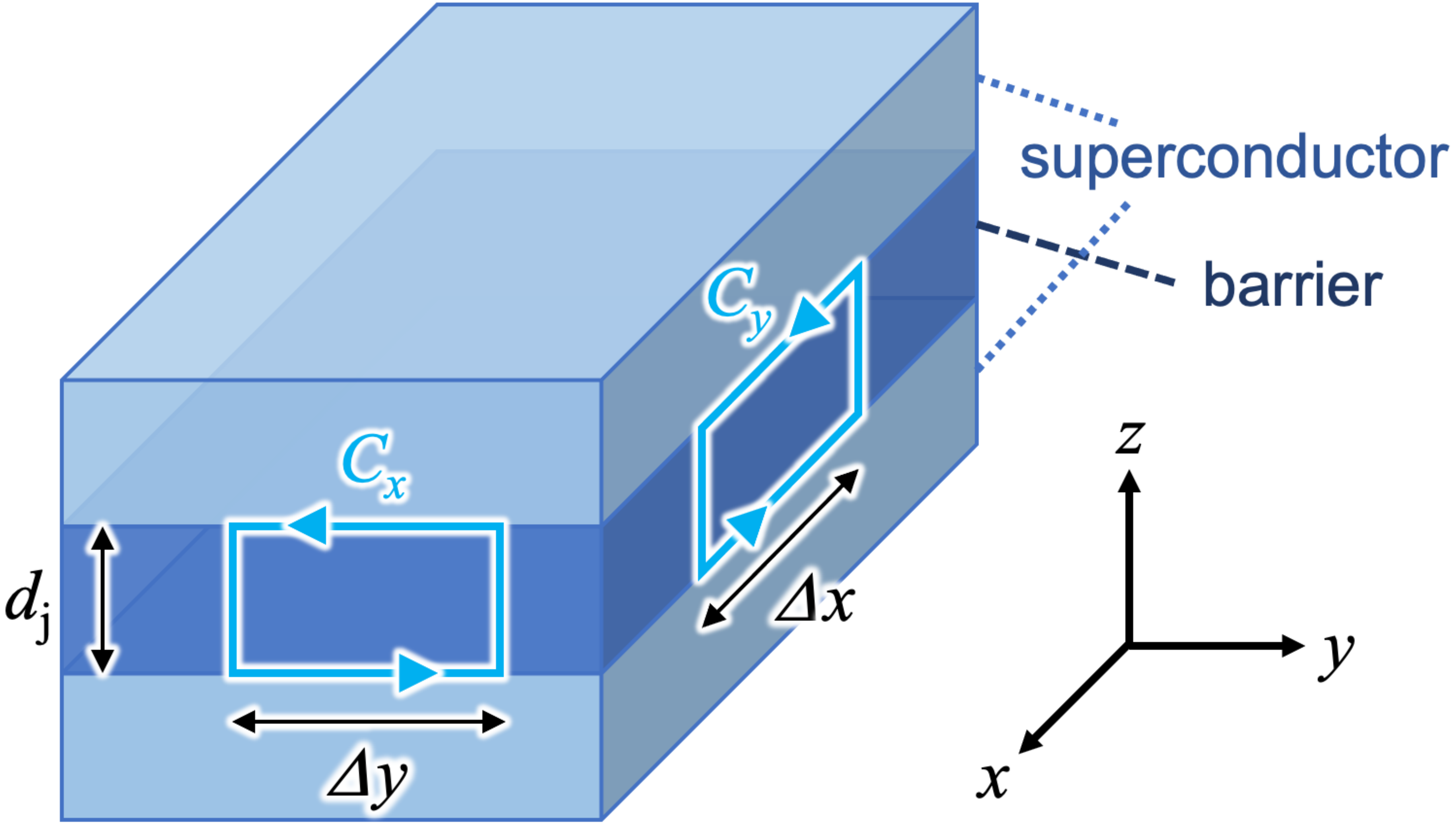}
\caption{%
Schematic of the Josephson-junction structure in which a barrier layer of thickness $d_{\rm j}$ located at $|z|<d_{\rm j}/2$ between superconducting layers at $|z|>d_{\rm j}/2$. 
$C_x$ and $C_y$ are closed rectangular contours of $\Delta y\times d_{\rm j}$ in the $yz$ plane and of $\Delta x\times d_{\rm j}$ in the $zx$ plane, respectively.}
\label{fig:JJ-structure}
\end{figure}

We consider a Josephson junction as shown in figure~\ref{fig:JJ-structure} in which a barrier layer of thickness $d_{\rm j}$ is located at $|z|<d_{\rm j}/2$ sandwiched between two superconducting layers. 

The supercurrent density associated with the order parameter $\psi= |\psi|\exp(i\varphi)$ based on Ginzburg-Landau theory is given by~\cite{Josephson_65,Barone-Paterno_82}  
\begin{equation}
	\bi{J}= \frac{1}{\mu_0\lambda^2} 
		\left( \frac{\phi_0}{2\pi}\nabla\varphi -\bi{A} \right) , 
\label{eq:Js-GL}
\end{equation}
where $\mu_0$ is the vacuum permeability, $\lambda$ is the London penetration depth, 
$\varphi$ is the phase, and $\bi{A}$ is the vector potential related to the magnetic induction $\bi{B}=\nabla\times\bi{A}$. 
We assume that suppression of $|\psi|$ from its equilibrium value is negligible. 

The gauge-invariant phase difference $\theta(x,y)$ at the junction, which is associated with the phase $\varphi(x,y,z)$ and the $z$ component of the vector potential $A_z(x,y,z)$, is defined by~\cite{Josephson_65,Barone-Paterno_82,Anderson_63} 
\begin{eqnarray}
	\theta(x,y) &= \varphi(x,y,+d_{\rm j}/2) -\varphi(x,y,-d_{\rm j}/2) 
% \nonumber\\ & \quad {}
	-\frac{2\pi}{\phi_0}\int_{-d_{\rm j}/2}^{+d_{\rm j}/2} A_z(x,y,z)dz . 
\label{eq:theta}
\end{eqnarray}
We calculate the contour integral of the vector potential along the rectangular contour $C_x$ which has vertices at $(x,y,\pm d_{\rm j}/2)$ and $(x,y+\Delta y,\pm d_{\rm j}/2)$ (figure~\ref{fig:JJ-structure}). 
Using Stokes theorem, we have 
\begin{equation}
	\oint_{C_x}\bi{A}\cdot d\bi{s} 
	= \int_y^{y+\Delta y}dy \int_{-d_{\rm j}/2}^{+d_{\rm j}/2}dz\ B_x 
	= B_x d_{\rm j} \Delta y . 
\label{eq:A-Cx_Hx}
\end{equation}
for a thin junction layer $d_{\rm j}\to 0$ and narrow contour width $\Delta y\to 0$. 
Using equations~\eref{eq:Js-GL} and \eref{eq:theta}, we also obtain the following expression: 
\begin{eqnarray}
	\oint_{C_x}\bi{A}\cdot d\bi{s} 
	&= \left[ {}-A_y(x,y,+d_{\rm j}/2) +A_y(x,y,-d_{\rm j}/2) \right] \Delta y 
		+\int_{-d_{\rm j}/2}^{+d_{\rm j}/2}dz 
		\left[ A_z(x,y+\Delta y,z) -A_z(x,y,z)\right]	
\nonumber\\
	&= {}-\frac{\phi_0}{2\pi}\frac{\partial\theta(x,y)}{\partial y}\Delta y 
		+\mu_0\lambda^2 \left[ J_y(x,y,+d_{\rm j}/2) 
		-J_y(x,y,-d_{\rm j}/2)\right]\Delta y . 
\label{eq:A-Cx_theta-Jy}
\end{eqnarray}
Equations~\eref{eq:A-Cx_Hx} and \eref{eq:A-Cx_theta-Jy} yield~\cite{Josephson_65,Barone-Paterno_82,Miller_85,Weihnacht_69,Gurevich_92} 
\begin{eqnarray}
	\frac{\phi_0}{2\pi}\frac{\partial\theta(x,y)}{\partial y} 
	&= {}-B_xd_{\rm j} +\mu_0\lambda^2 
		\left[ J_y(x,y,+d_{\rm j}/2) -J_y(x,y,-d_{\rm j}/2)\right] . 
\label{eq:dth/dy-Hx}
\end{eqnarray}
The contour integral along $C_y$ of $\bi{A}$ also yields a similar equation:  
 \begin{eqnarray}
	\frac{\phi_0}{2\pi}\frac{\partial\theta(x,y)}{\partial x} 
	&= {}+B_yd_{\rm j} +\mu_0\lambda^2 \left[ J_x(x,y,+d_{\rm j}/2) -J_x(x,y,-d_{\rm j}/2)\right] . 
\label{eq:dth/dx-Hy}
\end{eqnarray}

\subsection{Cross-type junctions with thin superconducting nanostrips} %********************
\begin{figure}[t]%*************
	\center\includegraphics[width=70mm]{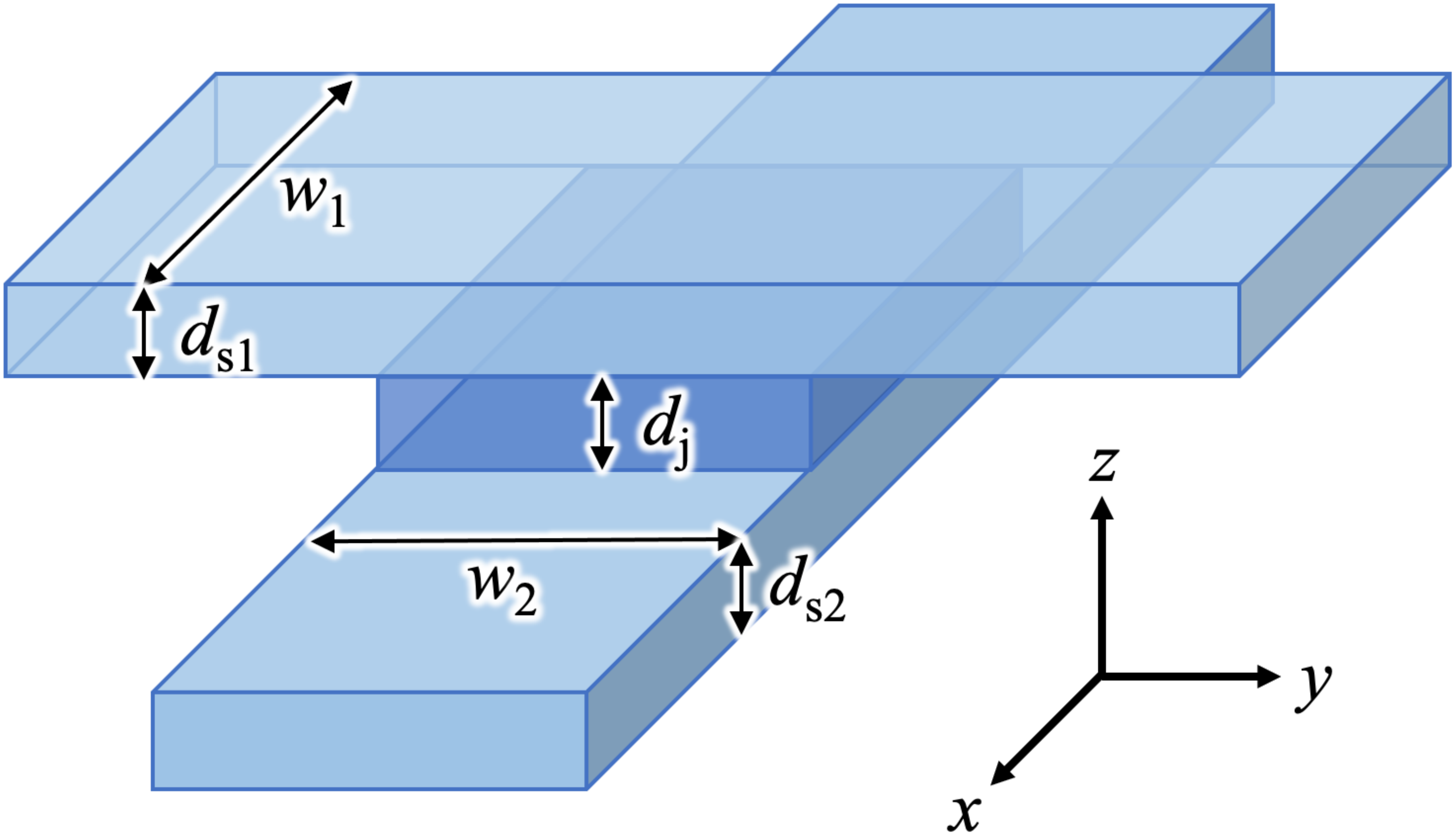}
\caption{%
Schematic of a cross-type Josephson junction with two superconducting nanostrips.  
The barrier layer is located at $|x|<w_1/2$, $|y|<w_2/2$, and $|z|<d_{\rm j}/2$. 
The top superconducting strip is located at $|x|<w_1/2$, $|y|<\infty$, and $d_{\rm j}/2<z<d_{\rm s1}+d_{\rm j}/2$, where $d_{\rm s1}<\lambda$ and $w_1<\lambda^2/d_{\rm s1}$. 
The bottom superconducting strip is located at $|x|<\infty$, $|y|<w_2/2$, and $-d_{\rm s2}-d_{\rm j}/2<z<-d_{\rm j}/2$, where $d_{\rm s2}<\lambda$ and $w_2<\lambda^2/d_{\rm s2}$. 
}
\label{fig:cross-strips}
\end{figure}

We now consider cross-type junctions with two superconducting nanostrips consisting of a top strip of thickness $d_{\rm s1}$ and width $w_1$ along the $y$ axis and a bottom strip of thickness $d_{\rm s2}$ and width $w_2$ along the $x$ axis, as shown in figure~\ref{fig:cross-strips}. 

We assume that the thicknesses of the strips are smaller than the London penetration depth, $d_{\rm s1}<\lambda$ and $d_{\rm s2}<\lambda$, and the widths are smaller than the Pearl length, $w_1<\lambda^2/d_{\rm s1}$ and $w_2<\lambda^2/d_{\rm s2}$, such that magnetic screening is weak in the superconducting nanostrips. 
The current density in thin and narrow superconducting strips is simply obtained by integrating the London equation, $\nabla\times\bi{J} =-\bi{B}/\mu_0\lambda^2 \simeq -\bi{H}/\lambda^2$, where $\bi{H}=(H_x,H_y,H_z)$ is the applied uniform magnetic field. 
The screening current density $\bi{J}_1$ in the top strip and $\bi{J}_2$ in the bottom strip are thus given by 
\begin{eqnarray}
	\bi{J}_1 &= \frac{1}{\lambda^2} \left[ -x H_z\hat{\bi{y}} 
		+\left( z-\frac{d_{\rm s1}+d_{\rm j}}{2} \right) 
		(-H_y\hat{\bi{x}}+H_x\hat{\bi{y}}) \right] , 
\label{eq:J1-upper}\\
	\bi{J}_2 &= \frac{1}{\lambda^2} \left[\, y H_z\hat{\bi{x}} 
		+\left( z+\frac{d_{\rm s2}+d_{\rm j}}{2} \right) 
		(-H_y\hat{\bi{x}}+H_x\hat{\bi{y}}) \right] . 
\label{eq:J2-lower}
\end{eqnarray}
Substitution of equations~\eref{eq:J1-upper} and \eref{eq:J2-lower} into equations~\eref{eq:dth/dy-Hx} and \eref{eq:dth/dx-Hy} yields
\begin{eqnarray}
	\frac{\phi_0}{2\pi\mu_0}\frac{\partial\theta(x,y)}{\partial x} 
	&= d_{\rm eff} H_y -y H_z , 
\label{eq:dth/dx-HyHz}\\
	\frac{\phi_0}{2\pi\mu_0}\frac{\partial\theta(x,y)}{\partial y} 
	&= -d_{\rm eff} H_x -x H_z ,  
\label{eq:dth/dy-HxHz}
\end{eqnarray}
where $d_{\rm eff}$ is the effective junction thickness for thin superconducting films as follows~\cite{Weihnacht_69}
\begin{equation}
	d_{\rm eff}= d_{\rm j}+(d_{\rm s1}+d_{\rm s2})/2 . 
\label{eq:d_eff}
\end{equation}
Integrating equations~\eref{eq:dth/dx-HyHz} and \eref{eq:dth/dy-HxHz}, we obtain the resulting distribution of the gauge-invariant phase difference, 
\begin{equation}
	\theta(x,y)= \theta_0 +\frac{2\pi\mu_0}{\phi_0} 
		\left[ d_{\rm eff}(x H_y -y H_x) -xy H_z \right] , 
\label{eq:theta-result}
\end{equation}
where $\theta_0$ is the integral constant. 
Note that $\theta$ is given as the sum of the parallel-field contribution ($\propto xH_y -yH_x$) and the perpendicular-field contribution ($\propto xy H_z$)~\cite{Miller_85},
and the simple formula of equation~\eref{eq:theta-result} is obtained for weak magnetic screening. 

Responding to the perpendicular field $H_z$, the screening current flowing in superconducting strips produces an effective parallel magnetic field in the junctions~\cite{Miller_85}, resulting in a modulation of $\theta$. 
Thus, the perpendicular field $H_z$ modulates $\theta$ through $J_x$ and $J_y$, as shown in the right-hand sides of equations~\eref{eq:dth/dy-Hx} and \eref{eq:dth/dx-Hy}.

\section{dc critical currents of cross-type Josephson junctions
\label{sec:Ic-cross-jj}} %************************************************************
\subsection{General expression for critical currents} %********************
The Josephson current density across the junction is $J_z= J_{\rm c}\sin\theta$, where $J_{\rm c}$ is the critical current density. 
The net current $I_z$ across the junction is given by 
\begin{eqnarray}
	I_z &= \int_{-w_1/2}^{w_1/2} dx \int_{-w_2/2}^{w_2/2} dy 
		\, J_{\rm c} \sin[\theta(x,y)] 
	= J_{\rm c} \,\mathrm{Im} \int_{-w_1/2}^{w_1/2} dx \int_{-w_2/2}^{w_2/2} dy 
		\,\exp\{i[\theta_0 +\theta_1(x,y)]\} ,  
\label{eq:Iz}
\end{eqnarray}
where equation~\eref{eq:theta-result} is rewritten as $\theta(x,y)= \theta_0 +\theta_1(x,y)$ and 
\begin{equation}
	\theta_1(x,y)= \frac{2x}{w_1}\beta -\frac{2y}{w_2}\alpha 
		-\frac{4xy}{w_1w_2}\gamma .  
\label{eq:theta-1}
\end{equation}
The parameters $\alpha$, $\beta$, and $\gamma$ in equation~\eref{eq:theta-1} are defined by 
\begin{eqnarray}
	\alpha =\pi{\mathit\Phi}_x/\phi_0 , \qquad
	&{\mathit\Phi}_x&= \mu_0H_xw_2 d_{\rm eff} ,
\label{eq:alpha}\\
	\beta =\pi{\mathit\Phi}_y/\phi_0 , 
	&{\mathit\Phi}_y&= \mu_0H_yw_1 d_{\rm eff} , 
\label{eq:beta}\\
	\gamma =\pi{\mathit\Phi}_z/2\phi_0 , 
	&{\mathit\Phi}_z&= \mu_0H_zw_1w_2 , 
\label{eq:gamma}
\end{eqnarray}
where ${\mathit\Phi}_x=\mu_0H_xS_x$, ${\mathit\Phi}_y=\mu_0H_yS_y$, and ${\mathit\Phi}_z=\mu_0H_zS_z$ are the magnetic fluxes linked in the areas $S_x=w_2d_{\rm eff}$, $S_y=w_1d_{\rm eff}$, and $S_z=w_1w_2$, respectively. 

The dc critical current $I_{\rm c}$ is obtained by maximizing $I_z$ with respect to $\theta_0$. 
Equation~\eref{eq:Iz} is rewritten in the form $I_z=\mathrm{Im}[f\exp(i\theta_0)]$, which has a maximum of $|f|$. 
We thus obtain $I_{\rm c}=J_{\rm c}|\int dx\int dy\exp(i\theta_1)|$~\cite{Josephson_65}. That is, 
\begin{equation}
	\frac{ I_{\rm c}({\mathit\Phi}_x,{\mathit\Phi}_y,{\mathit\Phi}_z) }{I_{\rm c0}}
	=\left| \mathcal{I} \left( \frac{\pi{\mathit\Phi}_x}{\phi_0}, 
		\frac{\pi{\mathit\Phi}_y}{\phi_0}, \frac{\pi{\mathit\Phi}_z}{2\phi_0} \right)\right| , 
\label{eq:Ic}
\end{equation}
where $I_{\rm c0}$ is the critical current at zero magnetic field, 
\begin{equation}
	I_{\rm c0}= I_{\rm c}(0,0,0)= J_{\rm c} w_1 w_2 ,
\label{eq:Ic0}
\end{equation}
and the complex-valued function $\mathcal{I}(\alpha,\beta,\gamma)$ is defined by 
\begin{equation}
	\mathcal{I}(\alpha,\beta,\gamma)= 
	\frac{1}{4}\int_{-1}^1dx' \int_{-1}^1dy' 
	\exp\left[ i(\alpha y' +\beta x' +\gamma x'y')\right] . 
\label{eq:cal-Ic}
\end{equation}
The integral variables in equation~\eref{eq:cal-Ic} are $x'=2x/w_1$ and $y'=-2y/w_2$. 
See \ref{app:cai-Ic_formulae} for details on $\mathcal{I}(\alpha,\beta,\gamma)$. 
The symmetry in $\mathcal{I}(\alpha,\beta,\gamma)$ shown in equation~\eref{eq:cal-Ic-symmetry} leads to symmetry in the critical current $I_{\rm c}\propto |\cal I|$, 
\begin{eqnarray}
	I_{\rm c}(-{\mathit\Phi}_x,{\mathit\Phi}_y,{\mathit\Phi}_z) 
	&= I_{\rm c}({\mathit\Phi}_x,-{\mathit\Phi}_y,{\mathit\Phi}_z) 
	=I_{\rm c}({\mathit\Phi}_x,{\mathit\Phi}_y,-{\mathit\Phi}_z) 
	= I_{\rm c}({\mathit\Phi}_x,{\mathit\Phi}_y,{\mathit\Phi}_z) . 
\label{eq:Ic-symmetry}
\end{eqnarray}
We thus consider only the case when ${\mathit\Phi}_x\geq 0$, ${\mathit\Phi}_y\geq 0$, and ${\mathit\Phi}_z\geq 0$.

\subsection{Critical currents in parallel or perpendicular fields} %********************
The $I_{\rm c}$ for parallel magnetic fields (i.e., ${\mathit\Phi}_z =0$) is obtained from equation~\eref{eq:cal-Ic-HxHy}, and is given by 
\begin{equation}
	\frac{I_{\rm c}({\mathit\Phi}_x,{\mathit\Phi}_y,0)}{I_{\rm c0}} 
	= \left| \frac{\sin(\pi{\mathit\Phi}_x/\phi_0)}{\pi{\mathit\Phi}_x/\phi_0} 
		\frac{\sin(\pi{\mathit\Phi}_y/\phi_0)}{\pi{\mathit\Phi}_y/\phi_0} \right| ,  
\label{eq:Ic-HxHy}
\end{equation}
which shows the standard Fraunhofer-like interference~\cite{Barone-Paterno_82}. 

The $I_{\rm c}$ for perpendicular magnetic fields (i.e., ${\mathit\Phi}_x ={\mathit\Phi}_y =0$) is obtained from equation~\eref{eq:cal-Ic-Hz}, and is given by~\cite{Miller_85} 
\begin{equation}
	\frac{I_{\rm c}(0,0,{\mathit\Phi}_z)}{I_{\rm c0}} 
	= \frac{\mathrm{Si}(\pi{\mathit\Phi}_z/2\phi_0)}{\pi{\mathit\Phi}_z/2\phi_0} ,  
\label{eq:Ic-Hz}
\end{equation}
where $\mathrm{Si}(z)=\int_0^z dt (\sin t)/t$ is the sine integral~\cite{Gradshtein_94}. 
Note that $I_{\rm c}(0,0,{\mathit\Phi}_z)$ monotonically decreases with increasing ${\mathit\Phi}_z$ and shows no interference patterns~\cite{Miller_85}. 
The contribution of the phase difference $\theta$ from the perpendicular field $H_z$ is proportional to $H_z$, as shown in equation~\eref{eq:theta-result}, and the modulation of $\theta$ due to $H_z$ results in the decrease of the critical currents as $I_{\rm c}\sim 1/H_z$ for large $H_z$~\cite{Miller_85,Monaco_08}.

\subsection{Critical currents in two-dimensional oblique fields} %********************
The $I_{\rm c}$ for ${\mathit\Phi}_y =0$ is obtained from equation~\eref{eq:cal-Ic-HxHz}, and is given by 
\begin{eqnarray}
	\frac{I_{\rm c}({\mathit\Phi}_x,0,{\mathit\Phi}_z)}{I_{\rm c0}} 
		&= \left| \frac{\mathrm{Si}[\pi({\mathit\Phi}_x+{\mathit\Phi}_z/2)/\phi_0] 
		-\mathrm{Si}[\pi({\mathit\Phi}_x-{\mathit\Phi}_z/2)/\phi_0]}{\pi{\mathit\Phi}_z/\phi_0} \right| . 
\label{eq:Ic-HxHz}
\end{eqnarray}
For ${\mathit\Phi}_x={\mathit\Phi}_z/2$ and ${\mathit\Phi}_y =0$ we have $I_{\rm c}({\mathit\Phi}_z/2,0,{\mathit\Phi}_z)=I_{\rm c}(0,0,2{\mathit\Phi}_z)$. 

\begin{figure}[b]%*************
	\center\includegraphics[width=80mm]{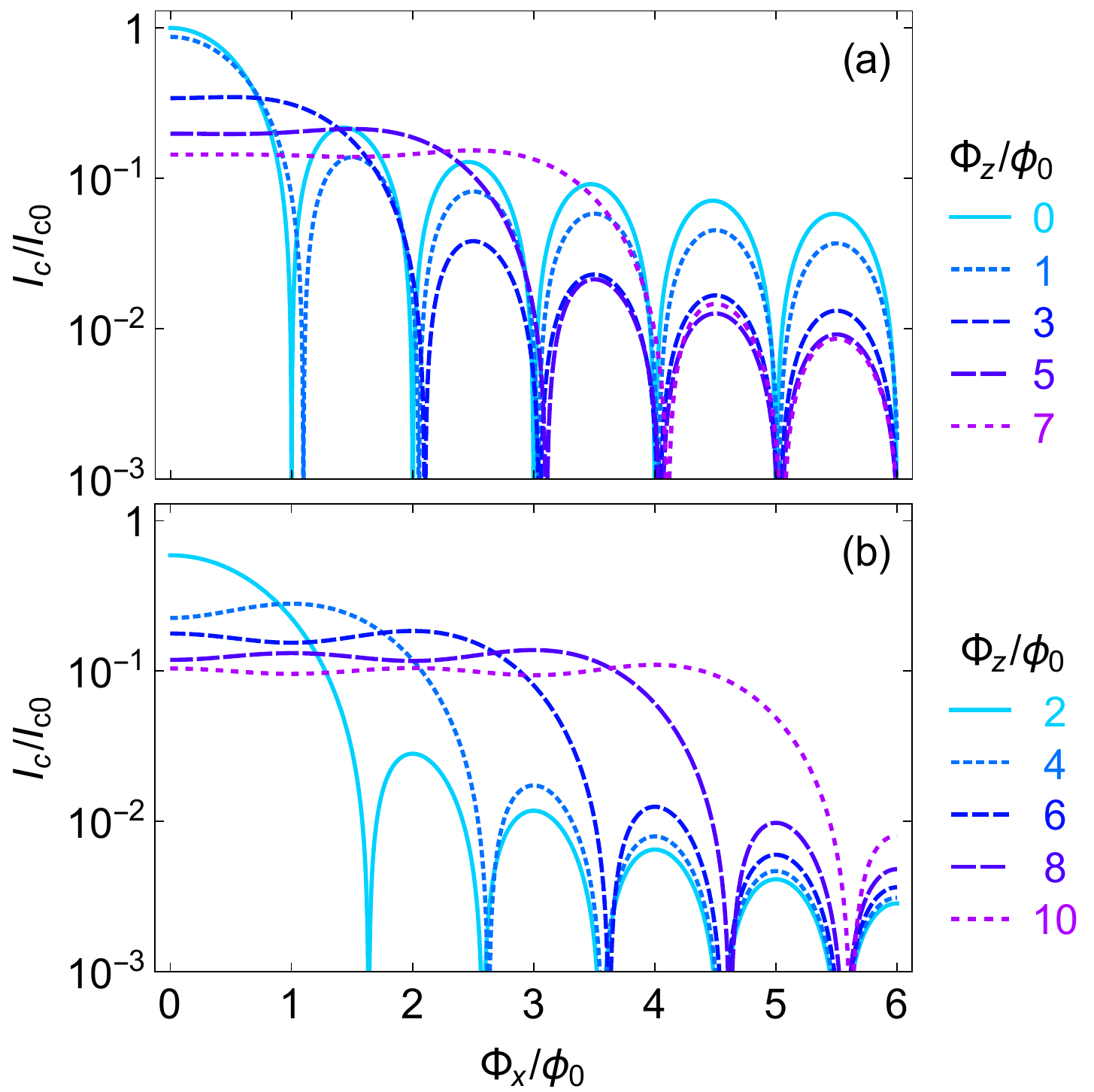} 
\caption{%
Critical current $I_{\rm c}({\mathit\Phi}_x,0,{\mathit\Phi}_z)/I_{\rm c0}$ vs parallel magnetic flux $\alpha/\pi={\mathit\Phi}_x/\phi_0$ for (a) $2\gamma/\pi={\mathit\Phi}_z/\phi_0= 0,\, 1,\, 3,\,5$, and $7$, and (b) ${\mathit\Phi}_z/\phi_0= 2,\, 4,\, 6,\,8$, and $10$.
}
\label{fig:Ic-Hx_Hz}
\end{figure}
\begin{figure}[bth]%*************
	\center\includegraphics[width=80mm]{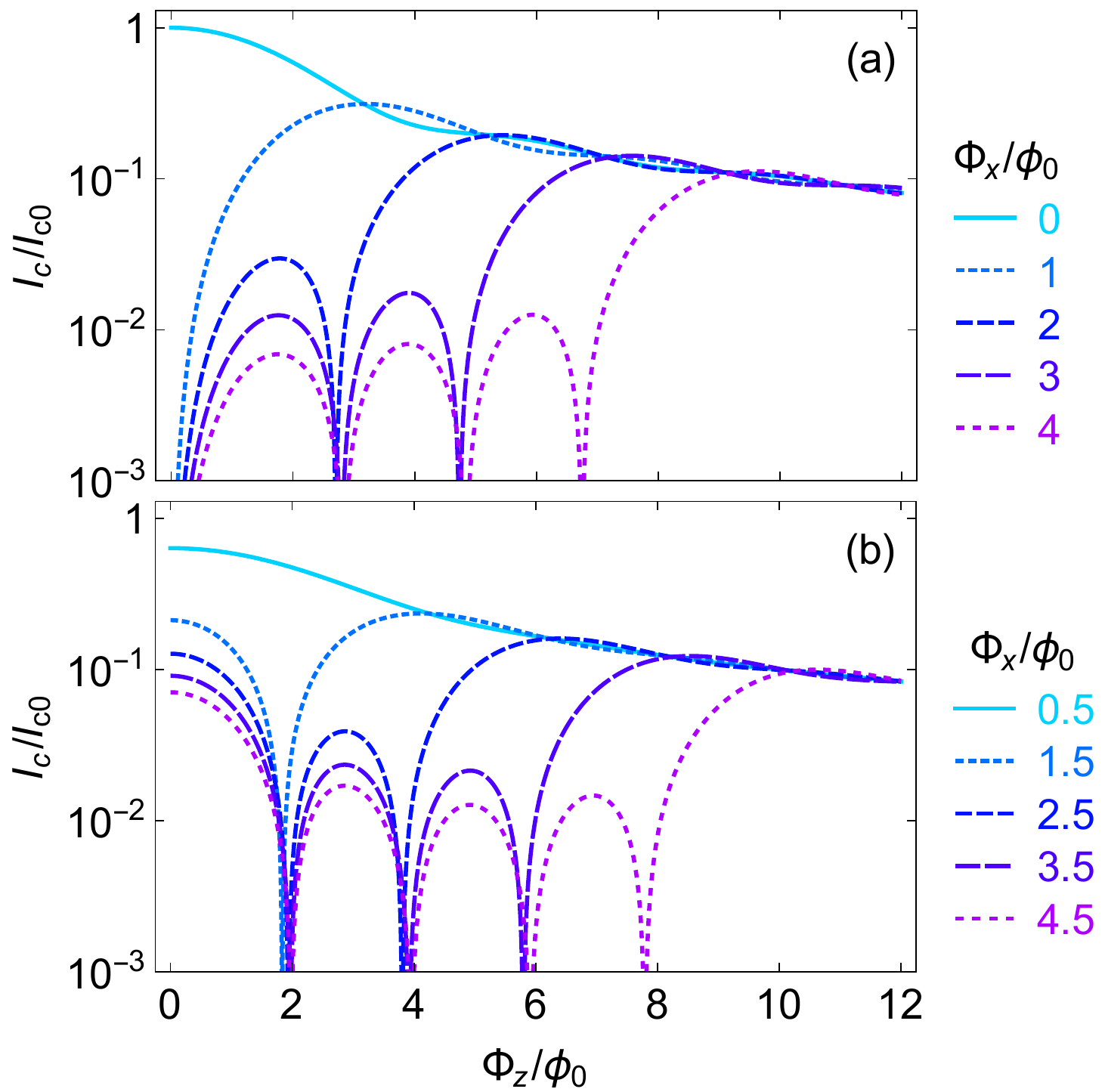} 
\caption{%
Critical current $I_{\rm c}({\mathit\Phi}_x,0,{\mathit\Phi}_z)/I_{\rm c0}$ vs perpendicular magnetic flux $2\gamma/\pi={\mathit\Phi}_z/\phi_0$ for (a) $\alpha/\pi={\mathit\Phi}_x/\phi_0= 0,\, 1,\, 2,\,3$, and $4$, and (b) ${\mathit\Phi}_x/\phi_0= 1/2,\,3/2,\,5/2,\,7/2$, and $9/2$.
}
\label{fig:Ic-Hz_Hx}
\end{figure}

Figures~\ref{fig:Ic-Hx_Hz} and \ref{fig:Ic-Hz_Hx} show the plots of $I_{\rm c}$ vs ${\mathit\Phi}_x$ and $I_{\rm c}$ vs ${\mathit\Phi}_z$, respectively, where $I_{\rm c}$ is obtained from equation~\eref{eq:Ic-HxHz}. 
The plot of $I_{\rm c}$ vs ${\mathit\Phi}_x$ for ${\mathit\Phi}_z=0$ in figure~\ref{fig:Ic-Hx_Hz}(a) corresponds to the Fraunhofer-like interference pattern given by equation~\eref{eq:Ic-HxHy} for ${\mathit\Phi}_y=0$. 
The plot of $I_{\rm c}$ vs ${\mathit\Phi}_z$ for ${\mathit\Phi}_x=0$ in figure~\ref{fig:Ic-Hz_Hx}(a) shows the monotonic dependence given by equation~\eref{eq:Ic-Hz}. 
Interference patterns appear when $|{\mathit\Phi}_z|< 2|{\mathit\Phi}_x|$. 
As shown in Figs.~\ref{fig:Ic-Hx_Hz}(a) and \ref{fig:Ic-Hz_Hx}(a), the $I_{\rm c}$ for ${\mathit\Phi}_z< 2{\mathit\Phi}_x$ is much smaller than $I_{\rm c0}$ when ${\mathit\Phi}_x/\phi_0\simeq 1,\,2,\,3,\,4,\,...$ and ${\mathit\Phi}_z/\phi_0\simeq 0,\,1,\,3,\,5,\,...$. 
As shown in Figs.~\ref{fig:Ic-Hx_Hz}(b) and \ref{fig:Ic-Hz_Hx}(b), the $I_{\rm c}$ for ${\mathit\Phi}_z< 2{\mathit\Phi}_x$ is much smaller than $I_{\rm c0}$ also when ${\mathit\Phi}_x/\phi_0\simeq 3/2,\,5/2,\,7/2,\,9/2,\,...$ and ${\mathit\Phi}_z/\phi_0\simeq 2,\,4,\,6,\,8,\,...$. 
The $I_{\rm c}$ with the magnetic interference is generally suppressed for $|{\mathit\Phi}_z|< 2|{\mathit\Phi}_x|$, while the $I_{\rm c}$ without the magnetic interference is relatively large for $|{\mathit\Phi}_z|> 2|{\mathit\Phi}_x|$. 
Therefore, the $I_{\rm c}$ temporarily increases with ${\mathit\Phi}_z$ at $|{\mathit\Phi}_z|\sim 2|{\mathit\Phi}_x|$, as seen in Fig.~\ref{fig:Ic-Hz_Hx}.

\begin{figure}[t]%*************
	\center\includegraphics[width=80mm]{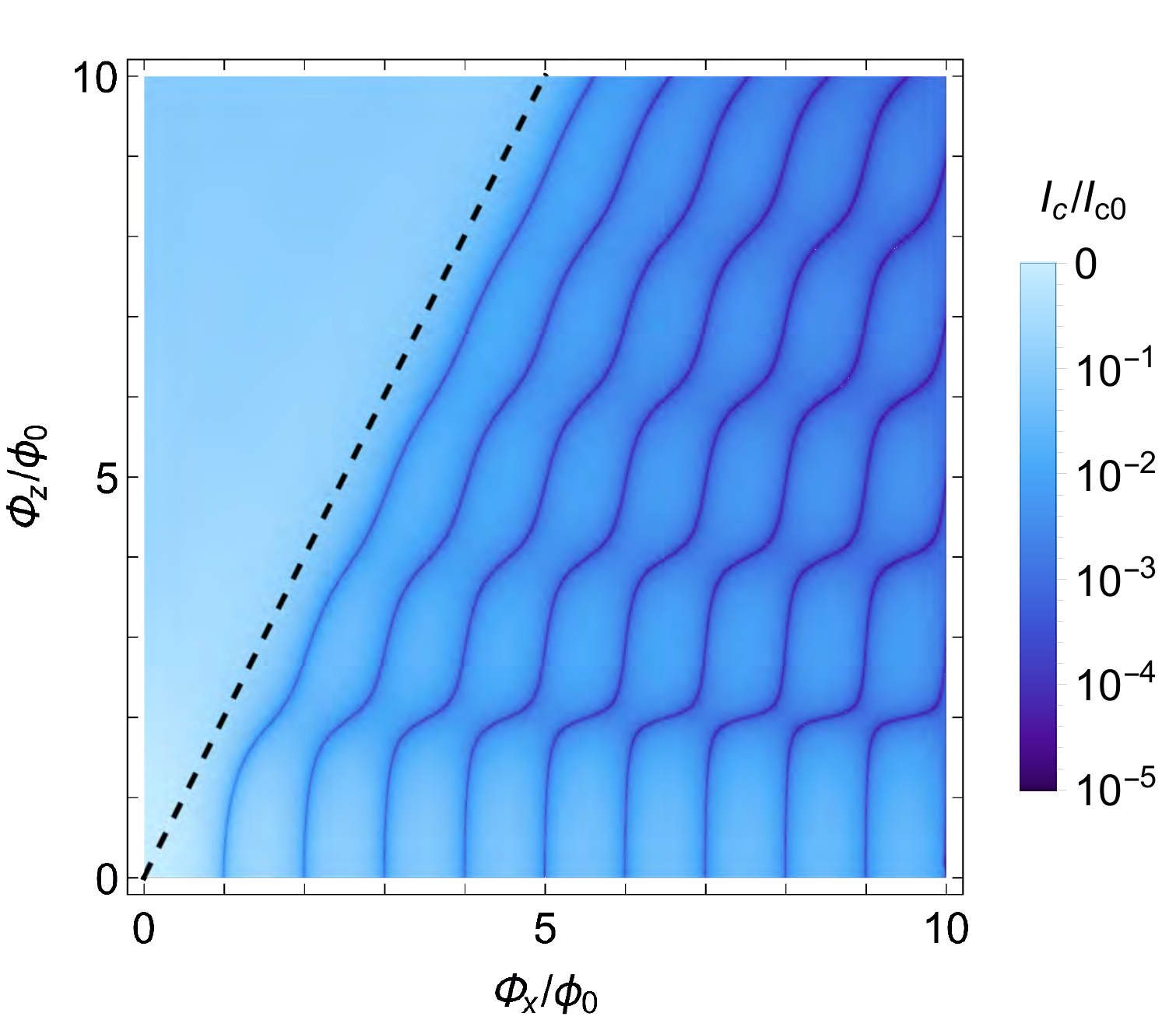} 
\caption{%
Logarithmic density plot of the critical current $I_{\rm c}({\mathit\Phi}_x,0,{\mathit\Phi}_z)/I_{\rm c0}$ as a function of the parallel $\alpha/\pi={\mathit\Phi}_x/\phi_0$ and perpendicular magnetic flux $2\gamma/\pi={\mathit\Phi}_z/\phi_0$.
The dashed line corresponds to ${\mathit\Phi}_z =2{\mathit\Phi}_x$. 
}
\label{fig:Ic-density_Hx-Hz}
\end{figure}

Figure~\ref{fig:Ic-density_Hx-Hz} shows density plots of $I_{\rm c}$ as a function of $({\mathit\Phi}_x,{\mathit\Phi}_z)$. 
For $|{\mathit\Phi}_z| < 2|{\mathit\Phi}_x|$ [i.e., the right-lower region below the dashed line in the $({\mathit\Phi}_x,{\mathit\Phi}_z)$ plane], the stepwise dark lines corresponding to $I_{\rm c}\ll I_{\rm c0}$ clearly demonstrate the interference patterns of $I_{\rm c}({\mathit\Phi}_x,0,{\mathit\Phi}_z)$, which are consistent with the behavior shown in Figs.~\ref{fig:Ic-Hx_Hz} and \ref{fig:Ic-Hz_Hx}. 

Substitution of $\alpha= (n+1/2)\pi$ or $\gamma= (m+1/2)\pi$ (where $n$ and $m$ are integers) in equation~\eref{eq:cal-Ic-largeHx-Hz} yields 
\begin{equation}
	\frac{I_{\rm c}({\mathit\Phi}_x,0,{\mathit\Phi}_z)}{I_{\rm c0}} 
	\sim \left| \frac{\sin(\pi{\mathit\Phi}_x/\phi_0)}{\pi{\mathit\Phi}_x/\phi_0} 
		\frac{\sin(\pi{\mathit\Phi}_z/2\phi_0)}{\pi{\mathit\Phi}_z/2\phi_0} \right| 
\label{eq:Ic_Hx-half_Hz-odd}
\end{equation}
for ${\mathit\Phi}_x/\phi_0 =n+1/2$ or ${\mathit\Phi}_z/\phi_0 =2m+1$. 
Equation~\eref{eq:Ic_Hx-half_Hz-odd} roughly explains the interference patterns shown in Figs.~\ref{fig:Ic-Hx_Hz}(a) and \ref{fig:Ic-Hz_Hx}(b). 
Substitution of $\alpha= n\pi$ or $\gamma= m\pi$ (where $n$ and $m$ are integers) into equation~\eref{eq:cal-Ic-largeHx-Hz} yields 
\begin{equation}
	\frac{I_{\rm c}({\mathit\Phi}_x,0,{\mathit\Phi}_z)}{I_{\rm c0}} 
	\sim \left| \frac{\cos(\pi{\mathit\Phi}_x/\phi_0)}{(\pi{\mathit\Phi}_x/\phi_0)^2} 
		\cos(\pi{\mathit\Phi}_z/2\phi_0) \right| 
\label{eq:Ic_Hx-integer_Hz-even}
\end{equation}
for ${\mathit\Phi}_x/\phi_0 =n$ or ${\mathit\Phi}_z/\phi_0 =2m$. 
Equation~\eref{eq:Ic_Hx-integer_Hz-even} roughly reproduces the interference patterns shown in Figs.~\ref{fig:Ic-Hx_Hz}(b) and \ref{fig:Ic-Hz_Hx}(a). 
Equation~\eref{eq:cal-Ic-Hx-largeHz} yields 
\begin{equation}
	\frac{I_{\rm c}({\mathit\Phi}_x,0,{\mathit\Phi}_z)}{I_{\rm c0}} 
	\sim \left| \frac{\phi_0}{{\mathit\Phi}_z} -\cos(\pi{\mathit\Phi}_x/\phi_0) 
		\frac{\cos(\pi{\mathit\Phi}_z/2\phi_0)}{(\pi{\mathit\Phi}_z/2\phi_0)^2} \right| , 
\label{eq:Ic_Hx-largeHz}
\end{equation}
which roughly explains the behavior of $I_{\rm c}$ without interference shown in Figs.~\ref{fig:Ic-Hx_Hz} and \ref{fig:Ic-Hz_Hx}. 

\subsection{Critical currents in three-dimensional oblique fields} %********************
\begin{figure*}[bth]%*************
	\center\includegraphics[width=\textwidth]{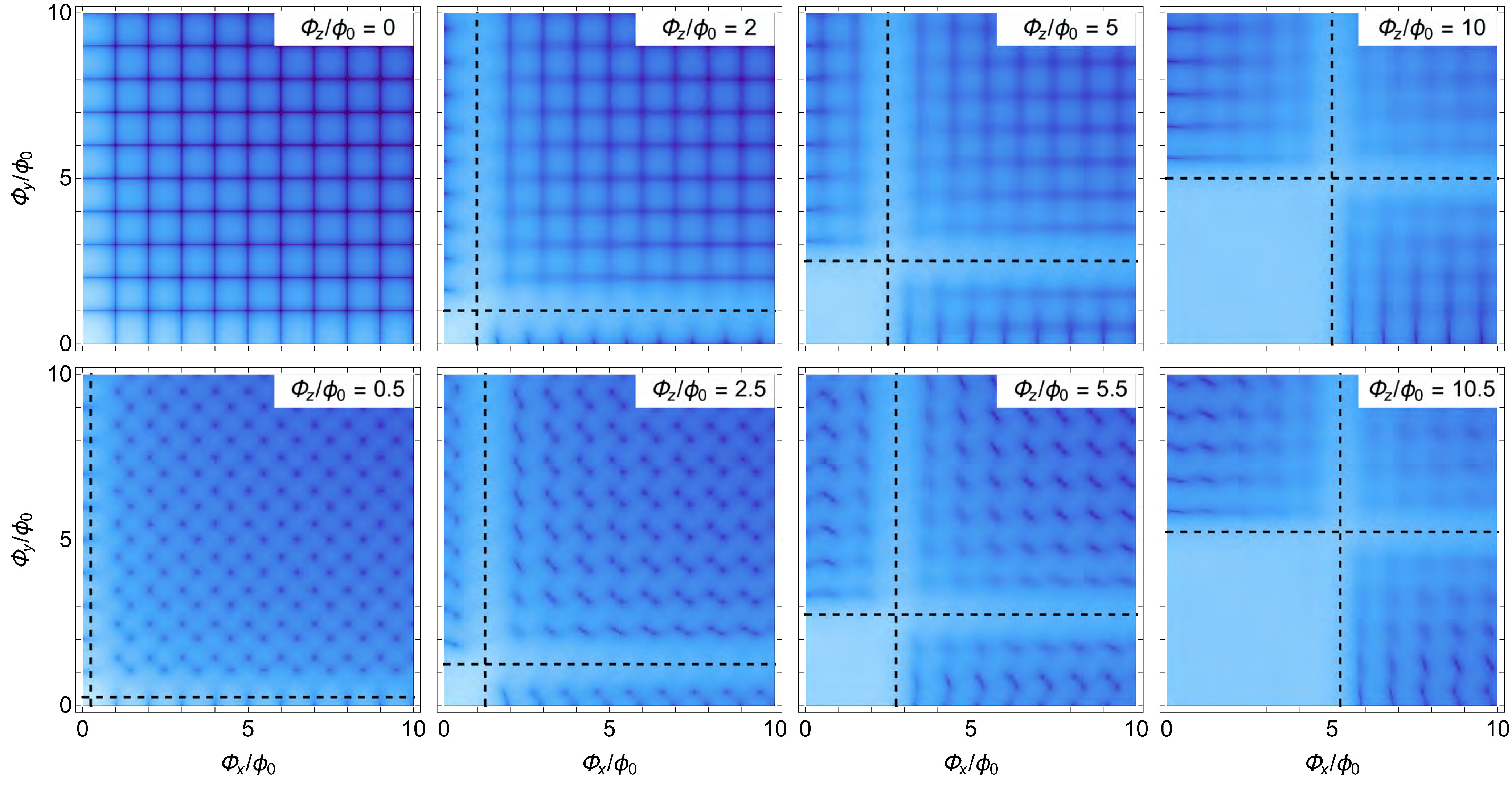} %180mm
\caption{%
Logarithmic density plots of the critical current $I_{\rm c}({\mathit\Phi}_x,{\mathit\Phi}_y,{\mathit\Phi}_z)/I_{\rm c0}$ as a function of $\alpha/\pi={\mathit\Phi}_x/\phi_0$ and $\beta/\pi={\mathit\Phi}_y/\phi_0$ when $2\gamma/\pi={\mathit\Phi}_z/\phi_0$ is integer (upper panels) and half integer (lower panels). 
The color scale in these plots is the same as in figure~\ref{fig:Ic-density_Hx-Hz}. 
Vertical and horizontal dashed lines correspond to ${\mathit\Phi}_z =2{\mathit\Phi}_x$ and ${\mathit\Phi}_z =2{\mathit\Phi}_y$, respectively.
}
\label{fig:Ic-Hx-Hy_Hz}
\end{figure*}

Figure~\ref{fig:Ic-Hx-Hy_Hz} shows density plots of $I_{\rm c}$ as a function of $({\mathit\Phi}_x,{\mathit\Phi}_y)$. 
The interference patterns of $I_{\rm c}$ disappear in the bright regions of $|{\mathit\Phi}_z|> 2\max(|{\mathit\Phi}_x|,|{\mathit\Phi}_y|)$. 
Dark lines or spots corresponding to the case where $I_{\rm c}\ll I_{\rm c0}$ show peculiar interference patterns with horizontal and vertical checkerboard patterns seen for integer ${\mathit\Phi}_z/\phi_0$ (upper panels of figure~\ref{fig:Ic-Hx-Hy_Hz}) and diagonal checkerboard patterns seen for half-integer ${\mathit\Phi}_z/\phi_0$ (lower panels of figure~\ref{fig:Ic-Hx-Hy_Hz}). 
These patterns are roughly reproduced by the simplified expression in equation~\eref{eq:Ic-HxHy-smallHz}. 

\begin{figure*}[b]%*************
	\center\includegraphics[width=\textwidth]{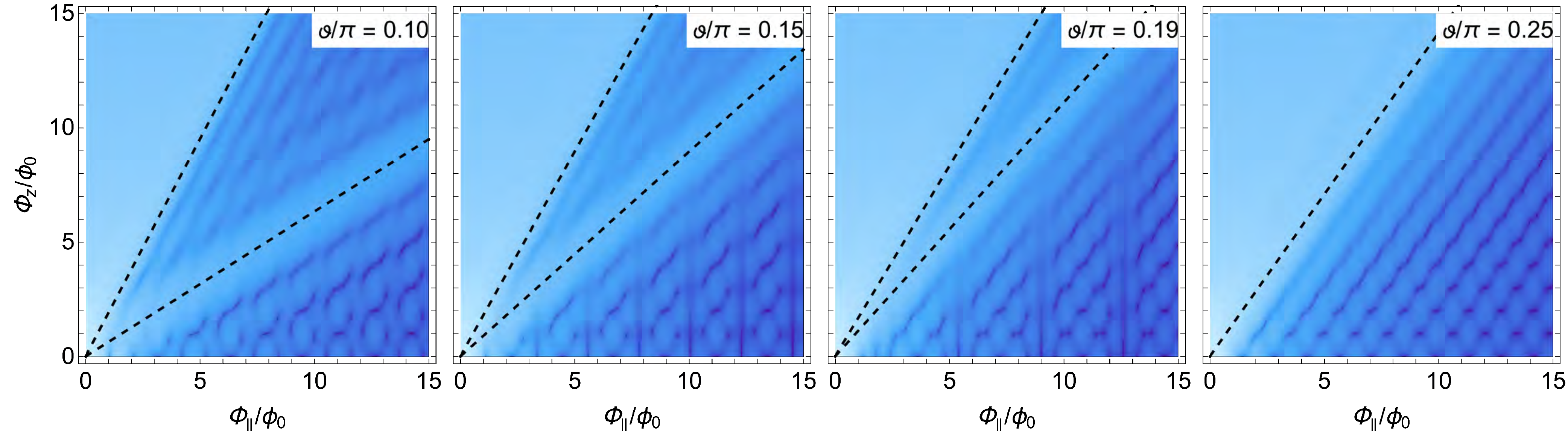} %180mm
\caption{%
Logarithmic density plots of the critical current $I_{\rm c}({\mathit\Phi}_{\parallel}\cos\vartheta,{\mathit\Phi}_{\parallel}\sin\vartheta,{\mathit\Phi}_z)/I_{\rm c0}$ as a function of ${\mathit\Phi}_{\parallel}\equiv({\mathit\Phi}_x^2+{\mathit\Phi}_y^2)^{1/2}$ and ${\mathit\Phi}_z$ for $\vartheta\equiv \arctan({\mathit\Phi}_y/{\mathit\Phi}_x)= 0.10\pi,\, 0.15\pi,\, 0.19\pi$, and $0.25\pi$. 
The color scale in these plots is the same as in figure~\ref{fig:Ic-density_Hx-Hz}. 
Upper and lower dashed lines correspond to ${\mathit\Phi}_z =2{\mathit\Phi}_x=2{\mathit\Phi}_{\parallel}\cos\vartheta$ and ${\mathit\Phi}_z =2{\mathit\Phi}_y=2{\mathit\Phi}_{\parallel}\sin\vartheta$, respectively.
}
\label{fig:Ic-Hxy-Hz_theta}
\end{figure*}

Figure~\ref{fig:Ic-Hxy-Hz_theta} shows the density plots of $I_{\rm c}$ as a function of $({\mathit\Phi}_{\parallel},{\mathit\Phi}_z)$, where ${\mathit\Phi}_{\parallel}\equiv ({\mathit\Phi}_x^2+{\mathit\Phi}_y^2)^{1/2}$. 
The parameter $\vartheta\equiv \arctan({\mathit\Phi}_y/{\mathit\Phi}_x)= \arctan[(w_1/w_2)(H_y/H_x)]$ corresponds to the field angle in the $xy$ plane when $w_1=w_2$. 
If we simply denote the $\vartheta$ dependence of the critical current as $I_{\rm c}(\vartheta)$, the $I_{\rm c}(\vartheta)$ has a periodicity expressed as $I_{\rm c}(\pi/2+\vartheta) =I_{\rm c}(\pi/2-\vartheta) =I_{\rm c}(\vartheta)$, as seen in Fig.~\ref{fig:Ic-Hx-Hy_Hz}. 
The $I_{\rm c}(\vartheta)$ for $0<\vartheta \leq\pi/4$ is plotted in Fig.~\ref{fig:Ic-Hxy-Hz_theta}. 
Noticeable $I_{\rm c}$-interference texture like weaving patterns, which can also be reproduced by equation~\eref{eq:Ic-HxHy-smallHz}, are seen in the region of ${\mathit\Phi}_z<2{\mathit\Phi}_{\parallel}\min(\cos\vartheta,\sin\vartheta)$. 

\section{Discussion and summary
\label{sec:summary}} %************************************************************
We considered the response of sandwich-type Josephson junctions in three-dimensional magnetic fields $\bi{H}=(H_x,H_y,H_z)$. 
It should be noted that the response to the perpendicular field $H_z$ is much more sensitive than that to the parallel magnetic field $H_{\parallel}$ if the junction dimensions in the $xy$ plane are much larger than the effective junction thickness $w_{\rm j}\gg d_{\rm eff}$, where $H_{\parallel}\equiv (H_x^2+H_y^2)^{1/2}$ and $w_{\rm j}\equiv \min(w_1,w_2)$. 
Even when $|H_z|\sim H_{\parallel}$, we have $|{\mathit\Phi}_z/{\mathit\Phi}_{\parallel}| \geq |H_z/H_{\parallel}| (w_{\rm j}/d_{\rm eff}) \gg 1$ from equations~\eref{eq:alpha}--\eref{eq:gamma}. 
Therefore, when the dependence of $I_{\rm c}$ on the field angle from the $xy$ plane  $\vartheta_{H}=\arctan(H_z/H_{\parallel})$ is investigated, the theoretical results presented in this paper may be observed in the narrow angle range of $|\vartheta_{H}| < d_{\rm eff}/w_{\rm j} \ll 1$. 
The present theory considers the case where the applied magnetic field is smaller than the lower critical field and no vortices are present in the junctions. In  perpendicular geometry, vortices are easily to penetrate into the junctions, and the perpendicular field $H_z$ should be small enough. 

The junction geometry and configuration of the superconducting films affect the interference behavior of $I_{\rm c}$. 
The $I_{\rm c}$ for overlap-type junctions in perpendicular fields $H_z$ show Fraunhofer-type interference patterns~\cite{Rosenstein_75,Hebard_75}, whereas $I_{\rm c}$ for cross-type junctions monotonically decrease with $H_z$ without interference~\cite{Miller_85}. 
Monaco \textit{et al}.~\cite{Monaco_08} numerically showed that interference patterns in $I_c$ vs $H_z$ strongly depend on the configuration of the superconducting strips. 
Although we consider only cross-type junctions in this paper, the $I_{\rm c}$ interference in oblique fields for other geometries (e.g., overlap-type junctions) is expected to show different patterns from our results. 

In this paper we assume that the critical current density $J_{\rm c}$ is homogeneous in the Josephson junctions, although inhomogeneous $J_{\rm c}$ affects the interference patterns of $I_{\rm c}$~\cite{Barone-Paterno_82,Dynes_71}. 
The inhomogeneous current distributions arising from the edge states in topological insulators~\cite{Lee_14,Hart_14,Pribiag_15} and in graphene~\cite{Allen_16} are investigated by analyzing the interference patterns of $I_{\rm c}$. 
It would be interesting to consider the effects of the inhomogeneous $J_{\rm c}(x,y)$ due to the edge states in topological materials upon the interference patterns of the sandwich-type junctions in three-dimensional magnetic fields. 

To summarize, we theoretically investigate the dc critical current $I_{\rm c}$ of cross-type Josephson junctions with thin and narrow superconducting strips exposed to three-dimensional magnetic fields $({\mathit\Phi}_x,{\mathit\Phi}_y,{\mathit\Phi}_z)$ for weak magnetic screening in the junctions and in the strips.
The standard Fraunhofer-type magnetic interference for parallel magnetic fields is strongly modulated by the perpendicular field, and the magnetic interference disappear when $|{\mathit\Phi}_z|> 2\max(|{\mathit\Phi}_x|,|{\mathit\Phi}_y|)$. 
Phase manipulation by the perpendicular field results in a variety of the magnetic interference of $I_{\rm c}$, and might be useful for developing novel Josephson-junction devices.

%\ack
\section*{Acknowledgments} %********************
I thank E. S. Otabe, T. Ueda, Y. Higashi, H. Asai, Y. Yamanashi, I. Kakeya, and M. Hidaka for stimulating discussions. 
This work has been supported by JSPS KAKENHI JP20K05314.

\appendix
\section{The complex-valued function $\mathcal{I}(\alpha,\beta,\gamma)$
\label{app:cai-Ic_formulae}} %************************************************************
We now present the mathematical formulae for the complex-valued critical current $\mathcal{I}(\alpha,\beta,\gamma)$ defined by equation~\eref{eq:cal-Ic}.
Because of the following symmetries, we need only consider the case when $\alpha\geq 0$, $\beta>0$, and $\gamma>0$:
\begin{equation}
	\mathcal{I}(-\alpha,\beta,\gamma)= \mathcal{I}(\alpha,-\beta,\gamma)=
	\mathcal{I}(\alpha,\beta,-\gamma)= \mathcal{I}^*(\alpha,\beta,\gamma) , 
\label{eq:cal-Ic-symmetry}
\end{equation}
where $\mathcal{I}^*$ is the complex conjugate of $\mathcal{I}$. 

The $\mathcal{I}$ for $\gamma=0$ and for $\alpha=\beta=0$ are respectively given by  
\begin{eqnarray}
	\mathcal{I}(\alpha,\beta,0)&= (\sin\alpha \sin\beta) /\alpha\beta , 
\label{eq:cal-Ic-HxHy}\\
	\mathcal{I}(0,0,\gamma)&= \mathrm{Si}(\gamma)/\gamma . 
\label{eq:cal-Ic-Hz}
\end{eqnarray}

The $\mathcal{I}$ for $\beta=0$ is 
\begin{equation}
	\mathcal{I}(\alpha,0,\gamma)= 
	\frac{1}{2\gamma} \left[ \mathrm{Si}(\alpha+\gamma) 
	-\mathrm{Si}(\alpha-\gamma) \right] . 
\label{eq:cal-Ic-HxHz}
\end{equation}
The asymptotic behavior of equation~\eref{eq:cal-Ic-HxHz} for $|\alpha|\gg |\gamma|$ is 
\begin{equation}
	\mathcal{I}(\alpha,0,\gamma) \sim 
	\frac{\sin\alpha \sin\gamma}{\alpha\gamma} 
	+\frac{\cos\alpha}{\alpha^2} \left(\cos\gamma -\frac{\sin\gamma}{\gamma}\right) ,  
\label{eq:cal-Ic-largeHx-Hz}
\end{equation}
and that for $|\alpha|\ll |\gamma|$ is 
\begin{equation}
	\mathcal{I}(\alpha,0,\gamma) \sim \frac{\pi}{2\gamma} 
		-\cos\alpha \frac{\cos\gamma}{\gamma^2} . 
\label{eq:cal-Ic-Hx-largeHz}
\end{equation}

Equation~\eref{eq:cal-Ic} is generally rewritten as 
\begin{equation}
	\mathcal{I}(\alpha,\beta,\gamma)= 
		\frac{i e^{-i\alpha\beta/\gamma}}{4\gamma} 
		\Bigl[ F(\eta_1)-F(\eta_2)-F(\eta_3)+F(\eta_4) \Bigr] , 
\label{eq:cal-Ic-F}
\end{equation}
where 
\begin{eqnarray}
	\eta_1 &= (\alpha+\gamma)(\beta+\gamma)/\gamma , 
\label{eq:eta-1}\\
	\eta_2 &= (\alpha-\gamma)(\beta+\gamma)/\gamma , 
\label{eq:eta-2}\\
	\eta_3 &= (\alpha+\gamma)(\beta-\gamma)/\gamma , 
\label{eq:eta-3}\\
	\eta_4 &= (\alpha-\gamma)(\beta-\gamma)/\gamma . 
\label{eq:eta-4}
\end{eqnarray}
The complex-valued function $F(z)$ in equation~\eref{eq:cal-Ic-F} is defined by 
\begin{equation}
	F(z)= \int_0^z \frac{1-e^{it}}{t}dt 
	= {}-i\,\mathrm{Si}(z) -\mathrm{Ci}(z) +\mathbf{C} +\ln z , 
\label{eq:F}
\end{equation}
where $\mathrm{Ci}(z)=-\int_z^{\infty}dt (\cos t)/t$ is the cosine integral and $\mathbf{C}=0.577...$ is Euler's constant~\cite{Gradshtein_94}. 
The general asymptotic behavior of equation~\eref{eq:cal-Ic-F} for small $\gamma$ [i.e., for $\max(|\alpha|,|\beta|)\gg |\gamma|$] is  
\begin{equation}
	\mathcal{I}(\alpha,\beta,\gamma) \sim \frac{1}{\alpha\beta} 
		( \sin\alpha\sin\beta\cos\gamma -i\cos\alpha\cos\beta\sin\gamma ) , 
\label{eq:Ic-HxHy-smallHz}
\end{equation}
and that for large $\gamma$ [i.e., for $\max(|\alpha|,|\beta|)\ll |\gamma|$] is 
\begin{eqnarray}
	\mathcal{I}(\alpha,\beta,\gamma) &\sim \frac{\pi}{2\gamma} 
		-\frac{i\pi\alpha\beta}{2\gamma^2} 
		-\frac{e^{i\alpha\beta/\gamma}}{\gamma^2} 
		( \cos\alpha\cos\beta\cos\gamma -i\sin\alpha\sin\beta\sin\gamma ) . 
\label{eq:Ic-HxHy-largeHz}
\end{eqnarray}

%\endrefs

\end{document}